\documentstyle[prl,aps,epsfig]{revtex}
\newcommand \be{\begin{equation}}
\newcommand \ee{\end{equation}}
\newcommand \beq{\begin{eqnarray}}
\newcommand \eeq{\end{eqnarray}}
\newcommand{\set}[2]{\newcommand{#1}{#2}}
\set{\pa}{\partial \over \partial\, }
\set{\leftvector}{\stackrel{\leftarrow}{\partial }}
\set{\rightvector}{\stackrel{\rightarrow}{\partial }}

\begin{document}

\twocolumn[\hsize\textwidth\columnwidth\hsize
           \csname @twocolumnfalse\endcsname

\title{Non-Linear Vibrations in Nuclei}
\author{C. Simenel and Ph. Chomaz}
\address{GANIL (DSM-CEA/IN2P3-CNRS), B.P.5027, F-14021 Caen cedex,
France}
\maketitle

\begin{abstract}
We have performed Time Dependent Hartree Fock (TDHF) calculations on
the non linear response of nuclei. We have shown that 
quadrupole (and dipole) motion produces monopole 
(and quadrupole) oscillations  in all atomic nuclei. 
We have shown that these findings can be interpreted as a 
large coupling between one and two phonon states 
leading to strong anharmonicities.
\end{abstract}

\vskip2pc
]

\section{Introduction}

Fifty years ago, it was discovered that atomic nuclei may enter in resonance
with electromagnetic fields\cite{to}. This Giant Dipole Resonance (GDR) has
been interpreted as the vibration of neutrons against protons. Since then,
other giant resonances (GR) have been predicted and observed, e.g. the
Monopole GR (GMR), an alternation of compression and decompression of the
nucleus, and the Quadrupole GR (GQR), an oscillation between a prolate and
an oblate shape. The proof of the vibrational nature of the GR came only few
years ago with the observation of the second vibrational quantum called the
two-phonon state\cite{cho1,aum}. While many properties of these states plead
in favor of an harmonic picture, striking experimental observations such as
an abnormally large excitation probability point to a strong coupling
between the different phonon states\cite{Vol1,Dasso}. This triggered a lot
of theoretical investigations but only very weak anharmonicities were found%
\cite{popo1,popo2,ccg2,lanza1,toya,brown} creating an important crisis in
our understanding of nuclear vibrations.

Recently, it has been proposed that a strong anharmonicity may come from
large residual interaction leading to the excitation of a GMR and a GQR on
top of any state \cite{Mumu}. This was a surprise especially in the
monopole case since it was generally believed that these couplings were small
because of cancellation effects between various diagrams \cite{wambach}. It
is shown in reference \cite{Mumu} that these cancellations between
3-particle 1-hole and 3-hole 1-particle matrix elements were very limited.
However, the approach of \cite{Mumu} even if it is fully microscopic do have
some drawbacks. It is based on boson mapping methods which may lead to
violation of the Pauli principle and mixing with the spurious states\cite
{Beaumel}. Therefore, an independent confirmation of these important
couplings leading to the excitation of a GMR and GQR on top of phonon states
is crucial.

In parallel looking to a completely different process, the
excitation of a GDR in fusion reactions, we have shown \cite{sim} that, in a
time dependent Hartree Fock (TDHF) approach, the dipole mode is non linearly
coupled with other collective modes such as in particular the vibration of
the density around a prolate shape. This work is also pointing in the
direction of anharmonic vibrations in nuclei but the particularities of the
fusion dynamics and of the composite system does not allow to draw
conclusions about the properties of the phonon built on the ground state.

In this work, we present the first realistic TDHF calculation \cite{bon} of
non-linear response to the collective vibrations showing that, indeed, the
one- to two- phonon coupling is a source of anharmonicities. We used
the TDHF approach \cite{har,foc,vau,bon,neg} which corresponds to an
independent propagation of individual particles in the self-consistent mean
field generated collectively. It does not incorporate the dissipation due to
two-body interaction \cite{gon,won,lac}, but takes into account one body
mechanisms such as Landau spreading and evaporation damping \cite{cho-Landau}%
. The quantal nature of the single particle dynamics is explicitly
preserved, which is crucial at low energy both because of shell effects and
of the wave dynamics. In its small amplitude limit TDHF is equivalent to the
Random Phase Approximation (RPA) which is the basic tool to understand the
collective response of nuclei in terms of independent phonons. However,
since the mean-field depends upon the actual excitation, TDHF is a non
linear theory and hence contains couplings between collective modes. This
point will be explicitly developed in the following. In fact TDHF is
optimized for the prediction of the average value of one body observables.
Through non-linearities, it takes into account the effects of the residual
interaction as soon as the considered phenomenon can be observed in the time
evolution of a one body observable. Of course, the absence of terms
explicitly taking into account the correlations is a limitation. In
particular, dampings and spreadings are neglected. As far as the time
dependent approaches are concerned, it would be important to extend the
present study to theories going beyond the one body limit such as extended
TDHF \cite{lac} which incorporate the effect of a ''collision term'' and
also, through the fluctuations associated with the considered dissipation,
the coherent coupling with phonon plus particle hole excitations. Even more
complete theories such as the time dependent density matrix approach\cite
{toya,casi}, which is known to reduce to the second RPA\ in its linearized
version, would be an interesting extension of the present work. Finally, one
should also try to apply the stochastic mean field approaches in particular
in its version which have been proved to be potentially an exact solution of
the many-body problem\cite{jui}. However, the analysis presented in section
3 clearly show that the non linear response in TDHF\ contains the couplings
between one and two phonon states coming from the 3-particle 1-hole and
1-particle 3-hole residual interaction.

In section 2 we demonstrate first that coupling between one and two
phonon states can be obtained through the evolution of average values of
one-body observables. In section 3, we show which part of the residual
interaction is taken into account in a TDHF approach. In section 4 we
present results demonstrating the importance of the non-linear excitation of
monopole and quadrupole modes on top of other collective vibrations. Finally
we will conclude in section 5. 

\section{\protect\smallskip Effect of couplings on one-body observables}

To understand how this coupling can be extracted from the one-body dynamics
let us consider the nonlinear coupling of a mode, $\mid \nu >$ with the GR$%
\mu $ built on top of it leading to the two phonon state $\mid \nu \mu >$.
The Hamiltonian can be written 
\[
{H}={H}_{0}+{V} 
\]
where ${H}_{0}$ corresponds to the harmonic (RPA) part for which $\mid \nu
>$ and $\mid \nu \mu >$ are eigenstates with energies $\omega _{\nu }$ and $%
\omega _{\nu \mu }=$ $\omega _{\nu }+$ $\omega _{\mu }$ while ${V}$ is the
residual interaction between phonons. For simplicity, let us introduce only
the non-linear coupling $v_{\mu }=<\nu \mid {V}\mid \nu \mu >$ which has
been proven to be the most important one\cite{Mumu}. At the first order in $%
\varepsilon =v_{\mu }/\omega _{\mu }$, this leads to the eigen states : 
\[
\mid \overline{\nu }>=({\mid \nu >-\varepsilon \mid \nu \mu >})/{\mathcal{N}}
\]
and 
\[
\mid \overline{\nu \mu }>=({\varepsilon \mid \nu >+\mid \nu \mu >})/{%
\mathcal{N}} 
\]
where ${\mathcal{N}}^{2}={{1+\varepsilon ^{2}}}$. A collective boost 
\[
\mid \psi (t=0)>=e^{-ik_{\nu }{Q}_{\nu }}\mid -> 
\]
inducing transitions between the ground state $\mid ->$ and the collective
state $\mid \nu >$ with the amplitude $q_{\nu }=<-\mid {Q}_{\nu }\mid \nu >$%
, leads to {\small 
\begin{equation}
\mid \psi (t)>\simeq \mid ->-ik_{\nu }q_{\nu }e^{-i\omega _{\nu }t}(\mid 
\overline{\nu }>-e^{-i\omega _{\mu }t}\varepsilon \mid \overline{\nu \mu }>).
\label{EQ:PSI-t-}
\end{equation}
} Then, $<{Q}_{\nu }>(t)$ is simply given by 
\begin{equation}
<{Q}_{\nu }>(t)\simeq -2k_{\nu }q_{\nu }^{2}\sin (\omega _{\nu }t).
\label{result2}
\end{equation}
This shows that the linear response to collective boost induces oscillation
of the collective moment at the collective frequency with an amplitude
proportional to the transition probability $q_{\nu }^{2}.$ If we now compute
the response to the operator ${Q}_{\mu }$ which is associated with the
excitation of the giant resonance $\mu ,$ it is not zero because of the
transitions between $\mid \overline{\nu }>$ and $\mid \overline{\nu \mu }>.$
Using Eq. (\ref{EQ:PSI-t-}) and assuming $<\nu \mu \mid {Q}_{\mu }\mid \nu
>=q_{\mu }$ too, we get at the lowest order in $k_{\nu }$ and $v/\omega
_{\mu }$ 
\begin{equation}
\,<{Q}_{\mu }>(t)=2k_{\nu }^{2}q_{\nu }^{2}q_{\mu }\frac{v_{\mu }}{\omega
_{\mu }}\left( \cos (\omega _{\mu }t)-1\right)  \label{result0}
\end{equation}
where the $-1$ term comes from higher order terms not explicitly written in
Eq. (\ref{EQ:PSI-t-}). This demonstrates that the induced moment $<{Q}_{\mu
}>\left( t\right) $ is quadratic in the collective boost amplitude as
expected from its non linear nature. Moreover, it oscillates at the
frequency of the coupled mode $\omega _{\mu }$ with an amplitude
proportional to the mixing coefficient $v_\mu/\omega _{\mu }$ and to the matrix
element $q_{\mu }.$ Finally, it should be noticed that $<{Q}_{\nu }>(t)$ and 
$<{Q}_{\mu }>(t)$ start in phase quadrature.

\section{Linear and non-linear response in TDHF}

The TDHF approach is built to describe the average values of one-body
observables. It propagates the evolution of the one-body density matrix 
\[
{\rho }_{ij}=<a_{j}^{+}a_{i}> 
\]
where $a_{i}^{+}$ is the operator creating a particle in the orbital $|i>$: 
\[
i\partial _{t}\rho =[h,{\rho }] 
\]
where ${h}$ is the self-consistent mean-field Hamiltonian linked to the mean
field energy $E$ by ${h}_{ij}=\partial E/\partial {\rho _{ji}}$ . We have
used the code of ref. \cite{kim} with $SGII$ \cite{sg} and $SLy4d$ \cite
{SLy4d} Skyrme interactions.

The RPA can be obtained by the linearization of the TDHF equation.
 Let us now go beyond the RPA by expanding the one-body density 
 $\rho$ up to the quadratic
terms in the collective boost strength $k_{\nu }$: 
\[
\rho =\rho ^{(0)}+k_{\nu }\rho ^{(1)}+k_{\nu }^{2}\rho ^{\left( 2\right) } .
\]
 $\rho ^{(0)}$, the static HF groundstate, defines the occupied states ($h$%
) ${\rho }^{(0)}=\sum_{h=1}^{A}\left| \varphi _{h}><\varphi _{h}\right| $,
the unoccupied states being the particle states ($p$). 
The condition on the one-body density $\rho$ due to the 
independent particle approximation made when deriving 
the TDHF equation leads to the constrain $\rho^{2}=\rho $. 
This imposes that $\rho ^{(1)}$ contains only $p$-$h$ 
components which directly provide the particle-particle 
and hole-hole elements of the quadratic term of 
the one-body density
\[
\rho _{pp^{\prime }}^{(2)}=\sum_{h}\rho _{ph}^{(1)}\rho _{hp^{\prime
}}^{(1)} 
\]
and 
\[
\rho _{hh^{\prime }}^{(2)}=-\sum_{p}\rho _{hp}^{(1)}\rho _{ph^{\prime
}}^{(1)}. 
\]

\subsection{\protect\smallskip TDHF and RPA}

The linear part of the TDHF equation leads to 
\[
i\partial _{t}{\rho }^{(1)}={{\mathcal{M}}}\rho ^{(1)}
\]
where 
\[
{\mathcal{M}}\cdot =[h^{(0)},\cdot ]+[\frac{\partial h}{\partial \rho }\cdot %
,\rho ^{\left( 0\right) }]
\]
is nothing but the RPA matrix acting only in the ph space. The RPA response
after a boost $e^{-ik_{\nu }Q_{\nu }}$ at time $t=0$ with $Q_{\nu }=q_{\nu
}O_{\nu }^{+}+h.c.$ where 
\[
\rho _{\nu }^{+}=[O_{\nu }^{+},\rho ^{\left( 0\right) }]
\]
is the RPA mode associated with the frequency $\omega _{\nu }$, is 
\[
\rho ^{(1)}=-q_{\nu }\rho _{\nu }^{+}ie^{-i\omega _{\nu }t}+h.c.\;.
\]
Using $\mathrm{Tr}O_{\nu }\rho _{\nu ^{\prime }}^{+}=\delta _{\nu \nu
^{\prime }}$, we get 
\[
<Q_{\nu }>=-2q_{\nu }^{2}k_{\nu }\sin \omega _{\nu }t
\]
which corresponds to Eq. (\ref{result2}) and explain why the RPA provides a
good approximation of $\omega _{\nu }$ and $q_{\nu }^{2}$.

\subsection{Quadratic response and phonon coupling}

If we now compute the quadratic response in the TDHF
approximation, we get for $\bar{\rho}^{(2)}$,  the $p-h$
component of the one-body density $\rho ^{(2)}$
%If we now compute the quadratic response in the TDHF approximation,
%we get for one-body 
%density $\bar{\rho}^{(2)},$ the $%
%p-h$ component of $\rho ^{(2)}$ 
\begin{equation}
i\partial _{t}\bar{\rho}^{(2)}={\mathcal{M}}\bar{\rho}^{(2)}+V^{d}+V^{e}+%
\delta V
\end{equation}
where the three sources of non linearities are the $p-h$ components of 
\begin{eqnarray*}
\delta V &=&\frac{1}{2}[\sum_{ijkl}\frac{\partial ^{2}h}{\partial \rho
_{ij}\partial \rho _{kl}}\rho _{ij}^{(1)}\rho _{kl}^{\left( 1\right) },\rho
^{\left( 0\right) }] \\
V^{d} &=&[\sum_{ij}\frac{\partial h}{\partial \rho _{ij}}\rho
_{ij}^{(1)},\rho ^{\left( 1\right) }] \\
V^{e} &=&[\sum_{ijk}\frac{\partial h}{\partial \rho _{ij}}\varepsilon
_{i}\rho _{ik}^{(1)}\rho _{kj}^{(1)},\rho ^{\left( 0\right) }]
\end{eqnarray*}
where $\varepsilon _{p}=1$ and $\varepsilon _{h}=-1$ and . 
It should be stressed that Eq. (4) is of course not the more 
general one but has the merit to be derived within the well 
defined framework of the time dependent mean field approximation.
The $p-h$
component of $\rho ^{(2)}$ can be expanded on the RPA basis 
\[
\bar{\rho}^{(2)}=q_{\nu }^{2}\sum_{\mu }z_{\mu }\rho _{\mu }^{+}+h.c. 
\]
where we have explicitly factorized the transition probability $q_{\nu }^{2}$%
. Using $\mathrm{Tr}O_{\mu }\rho _{\mu ^{\prime }}^{+}=\delta _{\mu \mu
^{\prime }}$, the evolution of $z_{\mu }$ can be isolated: 
\[
i\dot{z}_{\mu }=\omega _{\mu }z_{\mu }+v_{\mu }^{d}+v_{\mu }^{e}+\delta
v_{\mu } 
\]
with 
\[
v_{\mu }^{d}=\mathrm{Tr}O_{\mu }V^{d}/q_{\nu }^{2},v_{\mu }^{e}=\mathrm{Tr}%
O_{\mu }V^{e}/q_{\nu }^{2} 
\]
and 
\[
\delta v_{\mu }=\mathrm{Tr}O_{\mu }\delta V/q_{\nu }^{2}. 
\]
The time independent parts of $v_{\mu }^{d}+v_{\mu }^{e}$ $+\delta v_{\mu }$
, $v_{\mu }$, lead to 
\[
z_{\mu }=\frac{v_{\mu }}{\omega _{\mu }}(e^{-i\omega _{\mu }t}-1) 
\]
so that 
\[
<Q_{\mu }>=2k_{\nu }^{2}q_{\mu }\frac{q_{\nu }^{2}v_{\mu }}{\omega _{\mu }}%
\left( \cos \omega _{\mu }t-1\right) . 
\]
Comparing this result with the Eq. (\ref{result0}) shows that $v_{\mu }$ can
be interpreted as the residual interaction exciting, in TDHF, the mode $\mu $
on top of the phonon $\nu $.

\subsection{Link with the residual interaction}

To illustrate this coupling, let us compute the contribution coming from the
forward amplitudes $\rho _{\nu _{ph}}^{+}=O_{\nu _{ph}}^{+}=X_{ph}^{\nu }$
then $v_{\mu }^{d}$ contains two terms involving the $3p-1h$ and $3h-1p$
residual interaction, 
\begin{eqnarray}
v_{\mu }^{d}&=&\sum V_{p^{\prime }h^{\prime \prime };pp^{\prime \prime
}}X_{ph}^{\mu }X_{p^{\prime \prime }h^{\prime \prime }}^{\nu }X_{p^{\prime
}h}^{\nu } \nonumber \\
& &-\sum V_{hh^{\prime \prime };h^{\prime }p^{\prime \prime
}}X_{ph}^{\mu }X_{p^{\prime \prime }h^{\prime \prime }}^{\nu }X_{ph^{\prime
}}^{\nu } \nonumber
\end{eqnarray}
where we have introduced $V_{ik;jl}={\partial h_{ji}}/{\partial \rho _{kl}}$%
. Considering the second term $v_{\mu }^{e}$ we get also two components 
\begin{eqnarray}
v_{\mu }^{e}&=&\sum V_{p^{\prime }h^{\prime \prime };pp^{\prime \prime
}}X_{p^{\prime \prime }h^{\prime \prime }}^{\mu }X_{ph}^{\nu }X_{p^{\prime
}h}^{\nu } \nonumber \\
& &-\sum V_{hh^{\prime \prime };h^{\prime }p^{\prime \prime
}}X_{p^{\prime \prime }h^{\prime \prime }}^{\mu }X_{ph^{\prime }}^{\nu
}X_{ph}^{\nu } \nonumber
\end{eqnarray}
which are nothing but the exchange of $\mu $ and $\nu $ in $v_{\mu }^{d}$.
These four terms correspond exactly to the $X$ part of the phonon
interaction $\left\langle \mu \nu \right| V\left| \nu \right\rangle $
computed using boson mapping \cite{Mumu} except for a numerical factor which
actually depends upon the mapping used.

The $X$ part of $\delta v_{\mu }$ is $\sum \frac{\partial V_{h^{\prime
\prime }p^{\prime };p^{\prime \prime }h^{\prime }}}{\partial \rho _{ph}}%
X_{ph}^{\mu }X_{p^{\prime }h^{\prime }}^{\nu }X_{p^{\prime \prime }h^{\prime
\prime }}^{\nu }$. It results from the density dependence of the $p-h$
interaction defining the energy of the state $\nu $. In the simple case of a
linear density dependence, it can be interpreted as a contribution of a
three-body interaction inducing transitions from $1p-1h$ to $2p-2h$.

This analysis clearly shows that the time dependence in TDHF takes into
account the residual interaction. At the linear level, TDHF leads to the
RPA. Going to the quadratic response, TDHF takes into account the one- to
two-phonon coupling. The key point of the applicability of TDHF is the fact
that the studied phenomenon can be deduced from the time dependence of the
average value of a one-body observable.

\section{Results}

Let us now look at the TDHF results for the $^{40}Ca$ nucleus. We followed
the monopole, quadrupole and dipole response for three initial conditions:

\begin{itemize}
\item   A monopole boost using 
\[
{Q}_{0}=\frac{1}{\sqrt{4\pi }}\sum_{i}({r}_{i}^{2}-<{r}_{i}^{2}>(t=0)).
\]
because of the spherical symmetry, a monopole boost can only trigger
monopole modes. Therefore, we only observe $<{Q}_{0}>(t).$

\item   A quadrupole boost generated by 
\[
{Q}_{2}=\sum_{i}{r}_{i}^{2}Y_{0}^{2}\left( {\theta }_{i},{\varphi }%
_{i}\right) .
\]
The parity conservation forbids any dipole excitation when a quadrupole
velocity field is applied to a spherical nucleus. Conversely, breathing
modes (GMR) can be triggered by the quadrupole oscillation so that we do
follow both the quadrupole $<{Q}_{2}>(t)$ and the monopole $<{Q}_{0}>(t)$
responses.

\item   An isovector dipole boost induced by 
\[
{Q}_{D}=Z/A\sum_{n}z_{n}-N/A\sum_{p}z_{p}.
\]
This excitation can be both coupled to the quadrupole and monopole
oscillations so that we monitor the three moments, $<{Q}_{0}>(t)$, $<{Q}%
_{2}>(t)$ and $<{Q}_{D}>(t).$
\end{itemize}

%\smallskip

%%%\begin{center}
%%%\epsfig{figure=TDHF_PHONON_1.eps,height=9.cm}
%%%\end{center}

\begin{figure}[th]
    \begin{center}
    \epsfig{figure=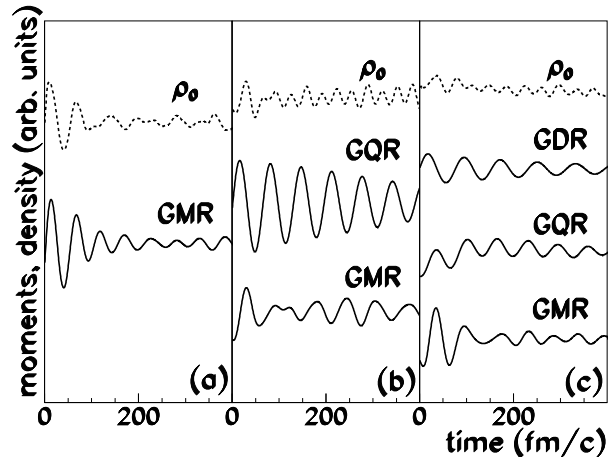,width=8.cm}
    \end{center}
\caption{Evolutions of the monopole, quadrupole and dipole moments (solid lines)
 and of the central density $\rho_0$ (dashed lines) as a
function of time for monopole (a), quadrupole (b) and dipole (c)  excitations
 in $^{40}Ca$.}
\label{fig1}
\end{figure}

In figure 1, we observe that the collective boost induces oscillations of
the associated moment as expected from the RPA (see Eq. (\ref{result2})).
They are only slightly damped in the GQR and GDR cases (fig. 1-b and 1-c respectively)
 while in the GMR case (fig. 1-a)
beatings, characteristic of a Landau damping, are observed. This means that
the dipole and quadrupole strengths are mostly concentrated in a single
resonance while the monopole one is fragmented.

Plotting in figure 2 the amplitude of the first oscillation $<{Q}_{\nu
}>_{max}$ as a function of $k_{\nu }$ confirms the linearity of this
response. Assuming that only one mode is excited which is a good
approximation for the GDR\ and GQR (Eq. (\ref{result2})) shows that the
transition probability, $q_{\nu }^{2},$ is $<{Q}_{\nu }>_{max}/2k_{\nu }.$
To get a deeper insight into the response we study the Fourier transform $%
F(\omega )$ of $<{Q}_{\nu }>(t)/k_{\nu }$ which is nothing but the RPA
strength when the velocity field $k_{\nu }$ is small enough to be in the
linear regime. We see in figure 3 that the dipole and quadrupole modes are
concentrated in a unique mode while the monopole is fragmented. However, the
various peaks are in the same energy region so that they can be approximated
by a single mode with a large Landau width. A detailed test of the
equivalence between the linear regime of TDHF and the RPA response can be
found in \cite{cho-Landau}.

If we now turn to the non linearities, we can observe the moments ${Q}_{\mu
} $ which are different from the operator ${Q}_{\nu }$ used for the exciting
boost. We see in figure 1 that, as expected from Eq. (\ref{result0}), this
non linear response follows a $\left( \cos (\omega _{\mu }t)-1\right)$
pattern oscillating with the frequency of the mode $\mu $ and not the one of
the initially excited collective state $\nu $. Moreover the amplitude of the
first oscillation (fig. 2) is as expected quadratic in the excitation
velocity $k_{\nu }$. In Fig. 1, one can see that large amplitude dipole (fig. 1-c) and
quadrupole (fig. 1-b) motion induces variations of the central density $\rho_0$. Since
the central density can be modified only by monopole states this imposes
that the large amplitude motion gets coupled with such breathing modes. In
the same way a large amplitude dipole oscillation induces a quadrupole
deformation of the nuclear potential and so gets coupled with the GQR. These
observations lead to the conclusion that we are in the presence of a
non-linear excitation of a giant resonance $\mu $ on top of the collective
motion $\nu $ initially excited through the collective boost ${Q}_{\nu }$.

%%%\begin{center}
%%%\epsfig{figure=TDHF_PHONON_2.eps,height=9.cm}
%%%\end{center}

\begin{figure}[th]
    \begin{center}
    \epsfig{figure=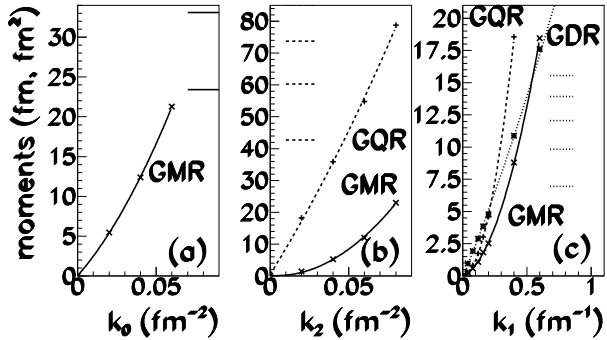,width=8.cm}
    \end{center}
\caption{Evolutions of the maximal oscillation amplitudes of $<{Q}_{0}>$ 
(solid line), $<{Q}_{2}>$ (dashed
line) and $<{Q}_{D}>$ (dotted line) as a function of the intensities of
the monopole (a), quadrupole (b), and dipole
excitations (c). The horizontal lines represent the average number of excited
phonons for each GR.}
\label{energy}
\end{figure}

The Fourier transform of $<{Q}_{\mu }>(t)$ associated with the excitation of 
${Q}_{\nu }$ are also presented in figure 3. Let us first start with the
quadrupole strength non-linearly excited by a dipole boost. This is a clear
indication that the observed state is indeed a GQR built on top of the GDR.
This is what is expected from Eq. (\ref{result0}) where only the frequency $%
\omega _{\mu }$ of the observed state appears. It should be notice that this
frequency is different from one of the underlying dipole motion. The
monopole case is more complex because of the presence of a strong Landau
spreading and it seems that the strengths of the various monopole states
depend upon the considered boost. This indicates that the coupling leading
to the excitation of an additional monopole state depends upon the
collective mode initially excited.

%%%\begin{center}
%%%\epsfig{figure=TDHF_PHONON_3.eps,height=9.cm}
%%%\end{center}

\begin{figure}[th]
    \begin{center}
    \epsfig{figure=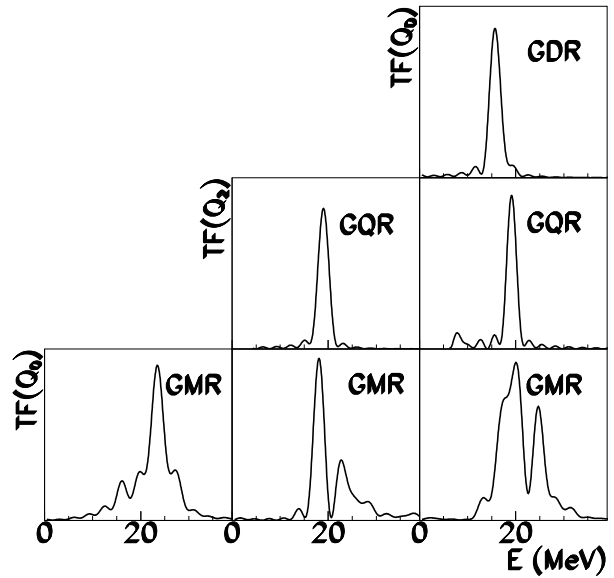,width=8.cm}
    \end{center}
\caption{Monopole, quadrupole and dipole spectra (the Fourier transforms are in arbitrary units) 
for a monopole (left), quadrupole (middle) and dipole excitation (right).}
\label{q0_q2}
\end{figure}

To estimate the magnitude of these non-linear couplings, we can first
convert the amplitude of the induced oscillations into a phonon number using
a coherent state picture. The horizontal lines in figure 2 represent the
amplitude of the oscillations associated with different number $<{n}_{\nu
}>=1,2,3,...$ of excited phonons: 
\[
<{Q}_{\nu }>_{max}^{2}=4<{n}_{\nu }>q_{\nu }^{2}.
\]
One can see that for a $k_{\nu }$ which corresponds to the excitation of one
phonon $\nu $ the number of phonons $\mu $ non-linearly excited is large.

Assuming for each multipolarity a unique state $|\mu >$ non linearly excited
one can use Eq. (\ref{result0}) to extract the residual interaction matrix
element $v_\mu$ between $|\nu >$ and $|\nu \mu >$ from the amplitudes of the
induced oscillations $<{Q}_{\mu }>_{max}$%
\begin{equation}
v_\mu={<{Q}_{\mu }>_{max}\omega _{\mu }}/{2k_{\nu }^{2}q_{\nu }^{2}q_{\mu }}.
\label{coupl}
\end{equation}
If the non-linear collective response is not concentrated in a unique state $%
|\mu >$ but corresponds to a set of states $\{|\mu _{i}>\}$ with $\omega
_{\mu _{i}}\approx \omega _{\mu }$, one can easily show that the extracted
coefficient $v_\mu$ is related to the individual ${v_\mu}_{i}=<\nu |V|\nu \mu _{i}>$
by the weighted sum 
\begin{equation}
v_\mu \approx \Sigma _{i}\sqrt{p_{\mu _{i}}}{v_\mu}_{i}  \label{weighted}
\end{equation}
where $p_{\mu _{i}}=q_{\mu _{i}}^{2}/{\Sigma _{j}q_{\mu _{j}}^{2}}$ and $%
q_{\mu _{i}}=<-|Q_{\mu }|\mu _{i}>$. %From this expression we seen that 
$v_\mu$ is in general higher than the individual ${v_\mu}_{i}$. For example, if the
collective response is equally distributed into $N$ states with identical
coupling matrix elements ${v_\mu}_{i}={\bar{v}_\mu}$ then $v_\mu=\sqrt{N}{\bar{v}_\mu}$.

\begin{table}[th]
    \begin{center}
{ 
\begin{tabular}{lccccc} \hline
{\footnotesize $|\nu>$} & {\footnotesize $\omega_{\nu}$} & 
{\footnotesize $q_{\nu}$} & {\footnotesize $S_1 $} & {\footnotesize $<\nu|%
{V}|\nu 0>$} & {\footnotesize $<\nu|{V}|\nu 2>$} \\ 
& {\footnotesize $(MeV)$} &  & $$ & {\footnotesize $(MeV)$}
& {\footnotesize $(MeV)$} \\ \hline
{\tiny $|0>_{^{40}Ca} $} & $22.9 $ & $11.6$  & $3090$  &        & $0 $
\\ 
{\tiny $|2>_{^{40}Ca} $} & $18.6 $ & $21.4$  & $9020$  & $-4.28 $ & $%
 $ \\ 
{\tiny $|D>_{^{40}Ca} $} & $17.2 $ & $3.47$  & $199 $  & $-4.58 $ & $%
-3.92 $ \\ \hline
{\tiny $|0>_{^{90}Zr} $} & $19.9 $ & $26.0$  & $12900$  &  & $0 
$ \\ 
{\tiny $|2>_{^{90}Zr} $} & $15.0 $ & $46.5$  & $35700$  & $-2.55 $ & 
 \\ 
{\tiny $|D>_{^{90}Zr} $} & $14.4 $ & $5.44$  & $442$  & $-1.60 $ & $-1.93 
$ \\ \hline 
{\tiny $|0>_{^{208}Pb}$} & $15.7 $ & $57.1$  & 
$51200$  &  & 0
 \\ 
{\tiny $|2>_{^{208}Pb}$} & $11.1 $ & $99.0$  & $128000$  & $-2.17 $ & 
 \\ 
{\tiny $|D>_{^{208}Pb}$} & $13.0 $ & $8.94$  & $877$  & $-2.40 $ & $-0.70 $
\\ \hline%
\end{tabular}
}
\end{center}

\caption{Energies, transition probabilities $q_\nu$, energy weighted sum-rules $S_1$ and coupling
coefficients of the GMR, GDR and GQR in the $^{40}Ca$, $%
^{90}Zr$ and $^{208}Pb$. $q_\nu$ and $S_1$ are expressed in $fm^2$
 and $MeV.fm^4$ for the GMR and GQR
 and in $fm$ and $MeV fm^2$ for the GDR respectively.}
\label{tdhf}
\end{table}

The results for the $^{40}Ca$, $^{90}Zr$ and $^{208}Pb$ are presented in
table 1. %******************************************************
The $\omega _{\nu }$ are computed from the time to reach the first maximum
of $<{Q}_{\nu }>(t)$. If the spreading of the observed mode is small, $%
\omega _{\nu }\approx {m_{2}}/{m_{1}}$ which should be a little higher than $%
{m_{1}}/{m_{0}}$ usually discussed. For the breathing mode, our results
agree with the values obtained in \cite{sg} (${m_{1}}/{m_{0}}=22.7$, $19.5$, 
$15.3MeV$ in $^{40}Ca$, $^{90}Zr$ and ${208}Pb$ respectively). These RPA
results as well as our $\omega _{\nu }$ are close to the averages which can
be computed from the most collective states reported in ref. \cite{Mumu}.
For the $^{40}Ca$, the ${m_{2}}/{m_{1}}$ for the monopole, quadrupole and
dipole states of ref. \cite{Mumu} are respectively $21.2$, $16.9$ and $18.5$ 
$MeV$ for $84\%$, $85\%$ and $66\%$ of the corresponding energy weighted sum
rules (EWSR). For ${208}Pb$ these values are $14.1$, $11.1$ and $13.6$ $MeV$
for $89\%$, $91\%$ and $80\%$ of the EWSR respectively. The relative sign of 
$v_\mu$ and $q_{\mu }$ is given by the early evolution of the moments described
by Eq. (5) and shown in figure \ref{fig1}. They appear to be all negative in
agreement with ref. \cite{Mumu}. The couplings $v_\mu$ are large of the order of
few MeV. From the quantitative point of view, the non-linear coupling
extracted from TDHF appears to be $50\%$ larger than the one reported in
reference \cite{Mumu}. This is a reasonable agreement since TDHF result is a
weighted sum of the individual couplings as shown in Eq. \ref{weighted}.
Summing the contributions of the different collective states considered in
ref. \cite{Mumu} reduces the difference between the reported values.
However, the phonon basis studied in ref. \cite{Mumu} being incomplete it is
expected that the TDHF results remains higher. It should be also noticed
that some difference can remain due to the approximations involved in the
different approaches as discussed in the quadratic response analysis.

In table 1 one can also see that the larger the nucleus the smaller the
coupling. This is in agreement with the fact that these couplings are
mediated by the surface. To control the robustness of our conclusion we have
performed a series of calculations using a different Skyrme force, the
recent SLy4d parametrization. For a $^{40}Ca$, this leads to a coupling
exciting the GMR on top of the GDR of $-4.01$ $MeV$ and of $-4.36$ $MeV$ on
top of the GQR. The quadrupole response during a dipole oscillation leads to
a residual interaction of $-3.98$ $MeV.$ Those results are very close to the
one reported in table 1.

\section{Conclusions}

In conclusion, we have shown with TDHF calculations that a non-linear
excitation of monopole and quadrupole should occur on top of any collective
motion in nuclei. These couplings can be interpreted in terms of a large
residual interaction which couples one-phonon and two-phonon states. 
These results show that large anharmonicities should be expected in the
collective motions in nuclei.

We thank Paul Bonche for providing his TDHF code and helpful discussions.
Comments and discussions with M. Andres, P.F. Bortignon, F. Catara, M.
Fallot, J. Frankland, O. Juillet, D. Lacroix and E. Lanza are acknowledged.

\end{document}